\documentclass[cameraready]{Interspeech}
\usepackage{amsmath,graphicx,hyperref}

\usepackage{subcaption}
\usepackage{booktabs}
\usepackage{graphicx}
\usepackage{pifont}
\usepackage{algorithm}
\usepackage{algpseudocode}
\title{Beyond One-Size-Fits-All: Personalized and Culturally Adaptive Emotional TTS via Interactive Optimization of Individual Emotion Perception Spaces}

\author[orcid=0009-0007-3139-0610]{Wangzixi}{Zhou}
\author[orcid=0000-0003-1560-2824]{Bagus Tris}{Atmaja}
\author[orcid=0000-0001-5509-8963]{Sakriani}{Sakti}

\address{
     Nara Institute of Science and Technology, Japan 
}

\email{zhou.wangzixi.az3@naist.ac.jp, bagus.tris@naist.ac.jp, ssakti@is.naist.jp
}

\keywords{emotional text-to-speech, personalization and cultural adaptation, emotion perception modeling}

\usepackage{comment}

\begin{document}

\maketitle


% need less 1000 characters
\begin{abstract}
The rise of conversational AI has increased interest in emotional Text-to-Speech (TTS). Most systems rely on discrete emotion labels, which fail to capture the nuanced nature of human affect. Recent models employ dimensional representations such as Russell’s arousal–valence (A–V) model, offering finer control. However, emotional perception varies across individuals and cultures, which may cause mismatches between modeled and perceived emotions. We propose a personalized and culturally adaptive emotional TTS framework that performs interactive optimization of individualized A–V perception spaces using an Interactive Genetic Algorithm. By adapting emotion representations to each listener, the system produces speech with more perceptually aligned emotional expression than models using averaged A–V values. Evaluations with Japanese, Chinese, and Indonesian participants highlight the importance of personalization and cultural adaptation for moving beyond one-size-fits-all emotional TTS.
%The rise of conversational AI has increased interest in emotional Text-to-Speech (TTS). Most systems rely on discrete emotion labels, which fail to capture the continuous and nuanced nature of human affect. Recent models using dimensional representations, such as Russell’s arousal–valence (A–V) model, offer finer control; however, emotion perception is subjective and culturally influenced, making generalized models inadequate. We propose a personalized and culturally adaptive emotional TTS framework that performs interactive optimization of individualized A–V perception spaces using an Interactive Genetic Algorithm. By adapting emotion representation to each listener, the system produces speech with more accurate and perceptually aligned emotional expression than model based on averaged A–V value. Cross-cultural evaluations in Japanese, Chinese, and Indonesian contexts demonstrate that personalization and cultural adaptation are essential to move beyond one-size-fits-all emotional TTS.
\end{abstract}

\vspace{-0.2cm}
\section{Introduction}
\vspace{-0.1cm}
Human emotion is inherently complex, shaped by subtle physiological, psychological, and social factors. Rather than existing as simple categorical states, emotional experiences lie on a continuum influenced by personality, gender, culture, and situational context \cite{fischer2004gender}. Capturing this variability is essential for speech synthesis systems that aim to interact naturally with humans, as emotion conveys intent, empathy, and engagement.

In affective computing, two widely adopted theoretical frameworks are the Emotion Wheel \cite{plutchik2013theories}, which organizes emotions into discrete categories (e.g., happiness, sadness, anger), and Russell’s Circumplex Model \cite{russell1980circumplex}, which represents emotions along continuous dimensions of arousal and valence. While state-of-the-art neural Text-to-Speech (TTS) systems \cite{mehta2024matcha,chen2025f5, shen2023naturalspeech, li2023styletts} have achieved high quality and naturalness, many emotional TTS approaches still rely on discrete style labels to represent emotion \cite{barakat2024deep, zhou2022emotional, diatlova2023emospeech, gao2025emo}. Such categorical representations limit the expression of subtle affective variations and often produce speech that sounds exaggerated or inflexible. To address this limitation, arousal–valence (A–V) representations have been explored in emotional TTS \cite{cho2024emosphere, cho2025emosphere++, liang2025ece} to enable finer emotional control and smoother transitions between affective states.

However, emotional perception is inherently subjective and culturally influenced. Models trained on averaged annotations—often derived from a single cultural population—implicitly assume a universal mapping between acoustic cues and perceived emotion. This assumption overlooks both individual variability and cross-cultural differences in emotional interpretation. Consequently, even when dimensional control is available, fixed A–V mappings may fail to align with a specific listener’s perception of emotion.

Recent large-scale preference learning approaches, such as reinforcement learning with human feedback (RLHF) \cite{ouyang2022training} and related alignment methods \cite{kaufmann2024survey}, have demonstrated success in aligning model behavior with human preferences through gradient-based retraining. However, these approaches typically target dataset-level alignment and require substantial preference data and model updates. In contrast, our objective is rapid per-user adaptation under extremely limited interactive feedback, without modifying the backbone acoustic model.

To address this challenge, we formulate emotional personalization as a low-dimensional perceptual optimization problem within the arousal–valence control space of neural TTS. We introduce a lightweight post-training personalization layer that refines emotion representations for each listener using direct preference feedback. As a practical instantiation of this interactive optimization process, we adopt an Interactive Genetic Algorithm (IGA) \cite{cho2002towards} to iteratively adjust A–V coordinates for target emotions. This design enables efficient adaptation with minimal user interaction, without retraining the pretrained TTS backbone or learning an explicit reward model.

Importantly, our approach personalizes the perceptual emotion mapping while keeping the high-dimensional acoustic representation fixed. By restricting adaptation to the low-dimensional emotion control space, the framework enables practical per-user customization while preserving model stability. Through this process, we obtain individualized variants of Russell’s circumplex model that better reflect each listener’s emotional interpretation.

\noindent Our key contributions are as follows:
\begin{itemize}
\item We propose a lightweight post-training personalization framework for emotional TTS that adapts arousal–valence representations to individual listeners under limited interactive feedback, instantiated using an Interactive Genetic Algorithm (IGA).
\item We design an emotion controller that maps personalized A–V coordinates to high-dimensional acoustic features, enabling fine-grained and perceptually aligned emotional expression.
\item We conduct a cross-cultural evaluation involving Japanese, Chinese, and Indonesian participants, demonstrating that both individual personalization and cultural adaptation significantly improve emotional alignment compared to one-size-fits-all models.
\end{itemize}

\noindent Speech samples from experiments are available on the demo page\footnote{https://37integer.github.io/Beyond-One-Size-Fits-All/}.

\begin{figure*}[t]  
    \centering

    \begin{minipage}{0.48\textwidth}
        \centering
        \includegraphics[width=\textwidth]{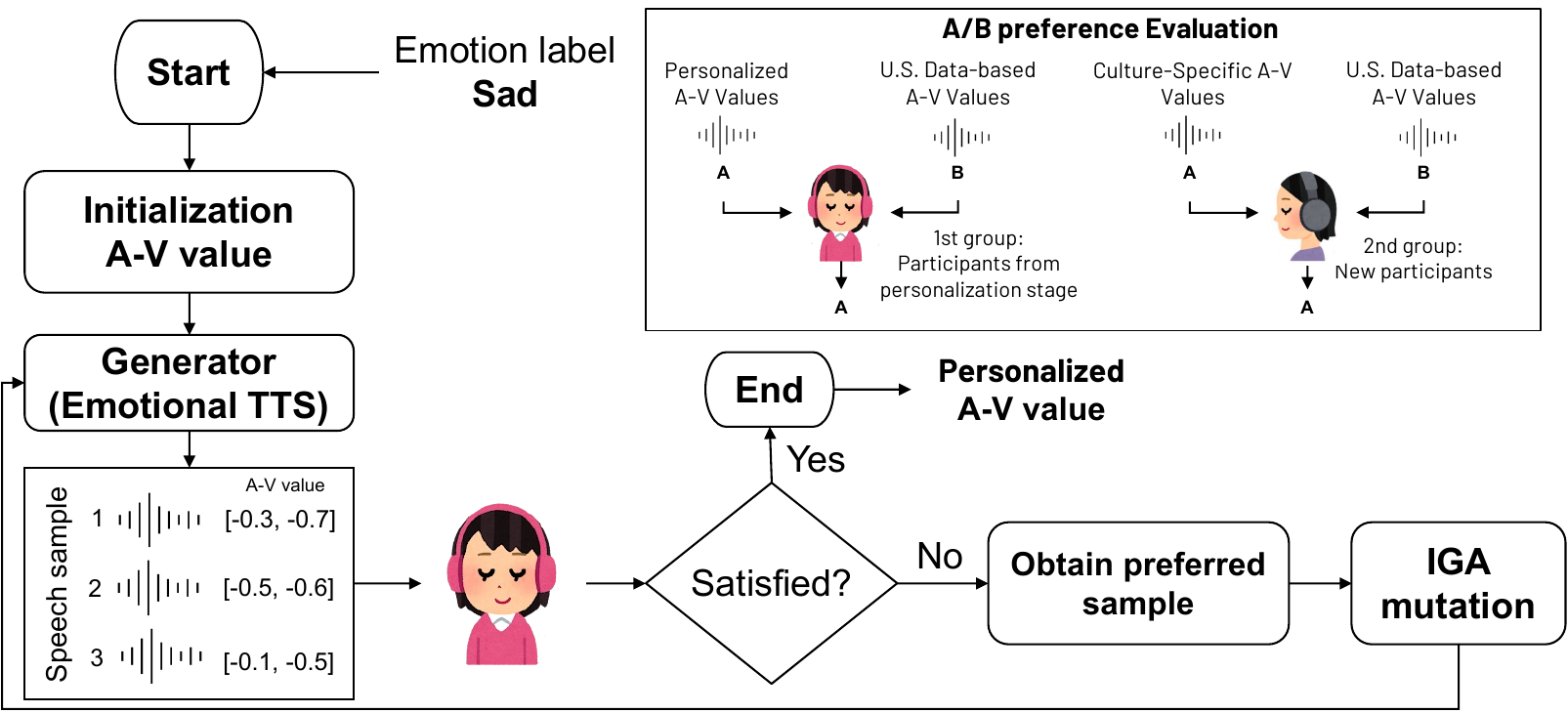}
        % \vspace{-0.1cm}
        \subcaption{Human-in-the-loop personalization with IGA}
        \label{fig:iga}
    \end{minipage}
    \hfill
    \begin{minipage}{0.48\textwidth}
        \centering
        \includegraphics[width=\textwidth]{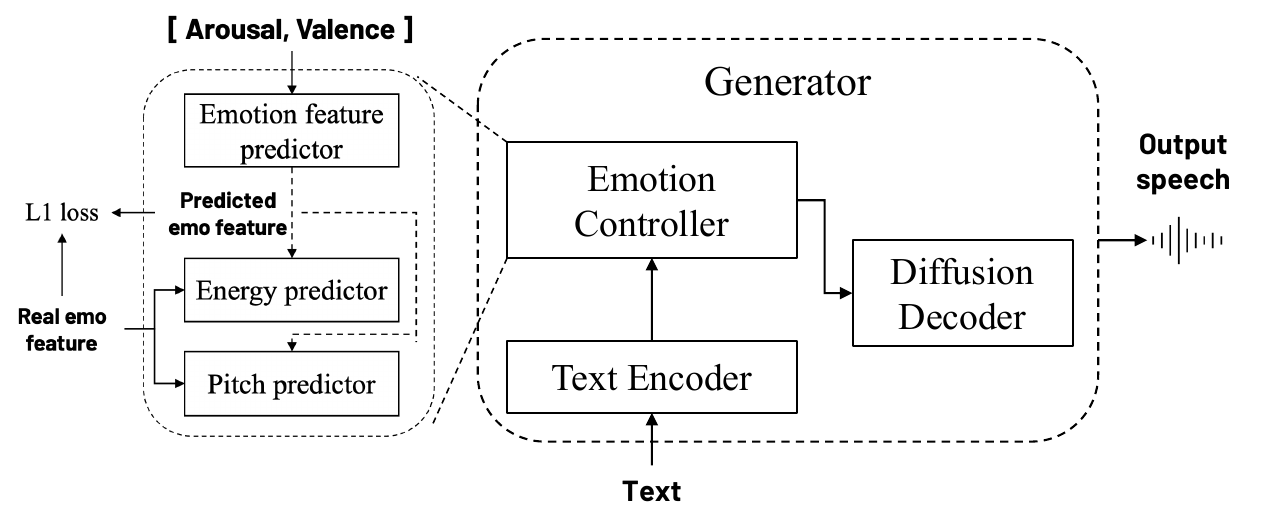}
        % \vspace{-0.1cm}
        \subcaption{Architecture of the emotional TTS generator}
        \label{fig:tts}
    \end{minipage}
    \vspace{-0.3cm}
    \caption{Overview of the proposed framework: (a) human-in-the-loop personalization using an Interactive Genetic Algorithm (IGA); (b) architecture of the emotional TTS generator with the proposed emotion controller.}
    \label{fig:architecture}
\end{figure*}

\vspace{-0.2cm}
\section{Related Work}
\vspace{-0.1cm}

\subsection{Controllable Emotional TTS}
\vspace{-0.1cm}

Emotional speech synthesis has evolved from categorical emotion representations (e.g., ``happy,'' ``sad'') \cite{barakat2024deep, zhou2022emotional, diatlova2023emospeech, gao2025emo} to dimensional emotion representations such as arousal, valence, and dominance \cite{cho2024emosphere, cho2025emosphere++, liang2025ece}. Dimensional representations enable smoother and more fine-grained affect modulation compared to discrete styles. However, both categorical and dimensional emotional TTS models are typically trained on averaged population-level annotations, implicitly assuming a universal mapping between acoustic cues and perceived emotion. Such an assumption overlooks individual and cultural variability in emotional interpretation, potentially leading to mismatches between synthesized emotion and listener perception.

\vspace{-0.1cm}
\subsection{Personalized Speech Synthesis}
\vspace{-0.1cm}

Personalization in TTS has primarily focused on speaker identity, voice timbre, and prosodic style modeling \cite{hsu2018hierarchical, shen2018natural, casanova2022yourtts}. These approaches enable voice cloning and flexible speaking style control but do not explicitly address personalization of emotional perception. Emotional expression in neural TTS is typically modeled using population-level representations, leaving individual and cultural differences in emotion interpretation largely unexplored. Extending personalization to emotional perception therefore remains an open challenge.

\vspace{-0.1cm}
\subsection{Preference-Guided Interactive Optimization}
\vspace{-0.1cm}

Preference-guided optimization incorporates human evaluation into the learning or search process, making it particularly suitable for subjective perceptual tasks. Large-scale alignment approaches such as reinforcement learning with human feedback (RLHF) \cite{ouyang2022training} and related techniques \cite{kaufmann2024survey} optimize model behavior through gradient-based retraining using large preference datasets, primarily targeting global model alignment.

In contrast, interactive optimization focuses on localized adaptation under limited feedback. The Interactive Genetic Algorithm (IGA) \cite{cho2002towards}, an extension of traditional genetic algorithms \cite{katoch2021review}, evolves candidate solutions based on user preference rather than explicit objective functions. IGA has been applied in visual design \cite{cho2002towards} and perceptual audio optimization \cite{fukumoto2020optimization}. While evolutionary search appeared in earlier unit concatenative TTS systems \cite{formiga2010evolutionary}, its use as a lightweight post-training perceptual adaptation mechanism for modern neural emotional TTS with continuous A--V control remains largely unexplored.

Taken together, these limitations motivate the need for a framework that directly adapts emotional representations to individual listeners through efficient interactive optimization.

%\begin{figure*}[t]  
%    \centering
%    \begin{minipage}{0.33\textwidth}
%        % \vspace*{0.15\textwidth}
%        \centering
%        \includegraphics[width=\textwidth]{china.png}
%        \subcaption{Chinese}
%        \label{fig:china}
%    \end{minipage}
%    \hfill
%    \begin{minipage}{0.33\textwidth}
%        \centering
%        \includegraphics[width=\textwidth]{indo.png}
%        \subcaption{Indonesian}
%        \label{fig:indo}
%    \end{minipage}
%    \hfill
%    \begin{minipage}{0.33\textwidth}
%        \centering
%        \includegraphics[width=\textwidth]{japan.png}
%        \subcaption{Japanese}
%        \label{fig:jap}
%    \end{minipage}
%    \vspace{-0.3cm}
%    \caption{Figure 2: A–V mappings from the personalization process for (a) Chinese, (b) Indonesian, and (c) Japanese participants. The $\bullet$ indicates participant-selected A–V values, and $\times$ indicates averaged reference dataset values.
%    }
%    \label{fig:culture}
%\end{figure*}

\vspace{-0.1cm}
\section{Proposed Framework}
\label{sec:method}
\vspace{-0.1cm}

We propose a personalized emotional TTS framework with human-in-the-loop preference learning. The overall architecture is illustrated in Fig.~\ref{fig:architecture} and consists of two components: (a) emotion preference learning that personalizes the arousal–valence (A–V) emotion space through user feedback, and (b) an emotion-controllable speech generator conditioned on A–V coordinates.

\vspace{-0.1cm}
\subsection{Emotion Preference Learning}
\vspace{-0.1cm}

As illustrated in Fig.~\ref{fig:iga}, we propose an emotion preference learning framework driven by an Interactive Genetic Algorithm (IGA). The primary objective is to personalize the mapping from discrete emotion categories to continuous A-V coordinates, aligning synthesized speech with each listener’s perceptual preferences. The optimization procedure is summarized in Algorithm~1. Given a target emotion (e.g., \textit{sad}), the system initiates an iterative optimization process: it generates a population of candidate A--V coordinates, synthesizes the corresponding speech samples, and dynamically refines these coordinates based on explicit user feedback until a highly satisfactory emotional expression is achieved.

% \vspace{-0.1cm}
\begin{algorithm}[htb]
\caption{Emotion Preference Learning via IGA}
\label{alg:iga}
\begin{algorithmic}[1]
\Require Target emotion $E$, generator $G$, population size $N$
\State Initialize candidate coordinates $\{e_i\}_{i=1}^{N}$
\State Set generation counter $g = 1$
\Repeat
 \For{$i=1$ to $N$}
    \State Synthesize speech samples $s_i = G(e_i)$ 
 \EndFor
     \State Present $\{s_i\}$ to user and obtain preferred sample $S_{pref}$
     \State Retain parent set $P = \{e_k \mid s_k \in S_{pref}\}$.
 \For{$i = |P| + 1$ to $N$} 
    \State Generate new candidate $\{e_k\}$ according to Eq. \ref{eq:crossover}
 \EndFor
    \State Update mutation strength $\mathcal{M}_{g+1}$ according to Eq. \ref{eq:decay}
    \State $g = g + 1$
\Until{user satisfaction}
\State \Return optimized personalized coordinate $e_k$
\end{algorithmic}
\end{algorithm}
% \vspace{-0.1cm}

During the preference learning, the user selects the samples that align with their perception of the target emotion. Based on this preference, the IGA evolves the new candidate set through a parent-based reproduction mechanism. The user-selected candidates will be directly preserved in the next generation. 
The algorithm utilizes multi-parent arithmetic crossover and uniform mutation. Specifically, given the preferred parent set $P$, a new child coordinate $e_k$ is generated through a random weighted combination of all parents in $P$:
\begin{equation}
e_k = \sum_{j=1}^{|P|} w_j p_j + \Delta, \label{eq:crossover}
\end{equation}
where $p_j \in P$, and $w_j \in [0,1]$ are random weights constrained by $\sum_{j=1}^{|P|} w_j = 1$. The term $\Delta \sim \mathcal{U}(-\mathcal{M}_g, \mathcal{M}_g)$ represents a random mutation perturbation applied independently to the arousal and valence dimensions.

To dynamically balance global exploration and local convergence, the mutation strength $\mathcal{M}_g$ decays multiplicatively over successive generations. The strength at generation $g$ is defined by the following piecewise decay function:
\begin{equation}
\mathcal{M}_g = \max(\mathcal{M}_{min}, \mathcal{M}_{g-1} \cdot \gamma), \label{eq:decay}
\end{equation}
where the initial mutation strength is set to $\mathcal{M}_1=0.20$, the decay rate is $\gamma=0.90$, and the lower bound is $\mathcal{M}_{min}=0.05$. This human-in-the-loop interaction typically converges within a few iterations once the listener's internal emotional criteria are met. Through this process, Russell’s circumplex model is adaptively reshaped for each listener, enabling emotional speech synthesis that reflects both individual and cultural perception patterns.

\vspace{-0.1cm}
\subsection{Emotional Speech Generator with Emotion Controller}
\vspace{-0.1cm}

%\vspace{-0.1cm}
%\subsection{Emotional Speech Generator}
%\vspace{-0.1cm}

The speech generator architecture is shown in Fig.~\ref{fig:tts}. We adopt Grad-TTS \cite{popov2021grad} as the backbone and use Russell’s circumplex model \cite{russell1980circumplex}, represented by arousal and valence (A–V), as the control interface for emotional speech synthesis. To achieve fine-grained emotional control, we introduce an Emotion Controller module that maps A–V coordinates to emotion-related acoustic features used by the generator.

%\vspace{-0.1cm}
%\subsection{Emotion Controller}
%\vspace{-0.1cm}

As shown in Fig.~\ref{fig:tts}, the Emotion Controller consists of three components: the proposed Emotion Feature Predictor, a Pitch Predictor, and an Energy Predictor (adapted from FastSpeech~2 \cite{ren2020fastspeech}). These modules jointly condition prosody on emotional cues and acoustic features, linking emotional state to low-level speech dynamics.

The Emotion Feature Predictor bridges the gap between the low-dimensional arousal–valence (A–V) space and high-dimensional emotion representations extracted from speech. It learns a mapping $f_{\theta}$ from an A–V coordinate $e=[a,v] \in \mathbb{R}^2$ to a latent emotion feature vector $h_{gt} \in \mathbb{R}^{D_{feat}}$.

Training targets $h_{gt}$ are obtained from a pre-trained Speech Emotion Recognition (SER) model \cite{wagner2023dawn}, which encodes high-level emotion-related acoustic features. Using these representations enables the model to learn emotion-related acoustic structure without additional manual annotation.

To enhance the representation capacity of the low-dimensional input, we apply Gaussian Fourier feature mapping:
\[
\gamma(e) = [\sin(2\pi eB), \cos(2\pi eB)] \in \mathbb{R}^{D_f},
\]
where $B \in \mathbb{R}^{2 \times (D_f/2)}$ is a fixed Gaussian matrix and $D_f$ is the Fourier feature dimension. The resulting feature vector $e_{ff} = \gamma(e)$ is fed into a four-layer MLP consisting of linear layers, SiLU activations, and dropout. A final linear projection maps the output to the target feature dimension $D_{feat}$, producing the predicted latent feature $h_{pred}$. The model is trained using an L1 loss between $h_{pred}$ and $h_{gt}$.

During training, pitch and energy predictors are conditioned on ground-truth emotion features. During inference, only the A–V coordinate is required, and the Emotion Feature Predictor generates the corresponding emotion representation used by the generator.

In our framework, the SER-derived feature space is trained prior to the IGA process and remains fixed, providing a stable acoustic prior for emotional speech synthesis. Personalization is achieved by adapting the A–V coordinates through human-in-the-loop preference learning, allowing users to reposition emotions according to subjective perception without retraining the acoustic model.

\vspace{-0.1cm}
\section{Experimental Setup}
% \vspace{-0.1cm}

\vspace{-0.1cm}
\subsection{Dataset and Training Setup}
\vspace{-0.1cm}

\noindent\textbf{Dataset:}
We constructed a 9-hour American English female emotional speech dataset (22.05 kHz) by combining EXPRESSO \cite{nguyen2023expresso}, EmoV-DB \cite{adigwe2018emotional}, and ESD \cite{zhou2022emotional}. Since these datasets do not provide continuous arousal–valence (A–V) annotations, we estimated A–V values using a pre-trained Speech Emotion Recognition (SER) model \cite{wagner2023dawn}.

\noindent\textbf{Model and Training:}
Acoustic features were 80-dimensional mel-spectrograms and HiFi-GAN \cite{Kong2020} was used as the vocoder. As a baseline, we implemented Grad-TTS with a simple A–V emotion embedding layer. Models were trained for 2000 epochs using Adam on a single NVIDIA RTX A6000 GPU. Default Grad-TTS hyperparameters were used unless specified.

\vspace{-0.1cm}
\subsection{Human-in-the-Loop Personalization}
\vspace{-0.1cm}

We recruited 30 participants from three Asian cultural groups: Chinese, Indonesian, and Japanese (10 per group, gender balanced). Seven emotions were used: angry, confused, sad, happy, disgust, sleepiness, and neutral.

Participants performed interactive personalization using the IGA loop for three sentences per emotion. In each round, participants selected the sample that best matched the target emotion. The process typically converged within three rounds, producing a personalized A–V mapping for each user. Cultural average A–V values were then computed for cross-cultural analysis.

\vspace{-0.1cm}
\subsection{Evaluation Protocol}
\vspace{-0.1cm}

\noindent\textbf{Objective Metrics:}
We evaluated 100 synthesized utterances using:

\begin{itemize}
\item \textbf{Word Error Rate (WER)} for speech intelligibility.
\item \textbf{Concordance Correlation Coefficient (CCC)} \cite{Atmaja2020b} to measure emotional similarity by comparing SER-predicted A–V values between generated and reference speech.
\end{itemize}

\noindent\textbf{Subjective Metrics:}
Listening tests involved 30 participants.

\begin{itemize}
\item \textbf{MOS:} Participants rated the naturalness of 20 utterances on a five-point scale.
\item \textbf{A/B preference:} Two groups were used. The first group (15 participants from the personalization stage) compared personalized A–V speech with the baseline using averaged U.S. dataset A–V values. The second group (15 new participants) evaluated culturally adapted speech using Japanese, Chinese, and Indonesian A–V averages against the same baseline. All tests were conducted blindly.
\end{itemize}

\begin{figure*}[t]  
    \centering
    \begin{minipage}{0.33\textwidth}
        % \vspace*{0.15\textwidth}
        \centering
        \includegraphics[width=\textwidth]{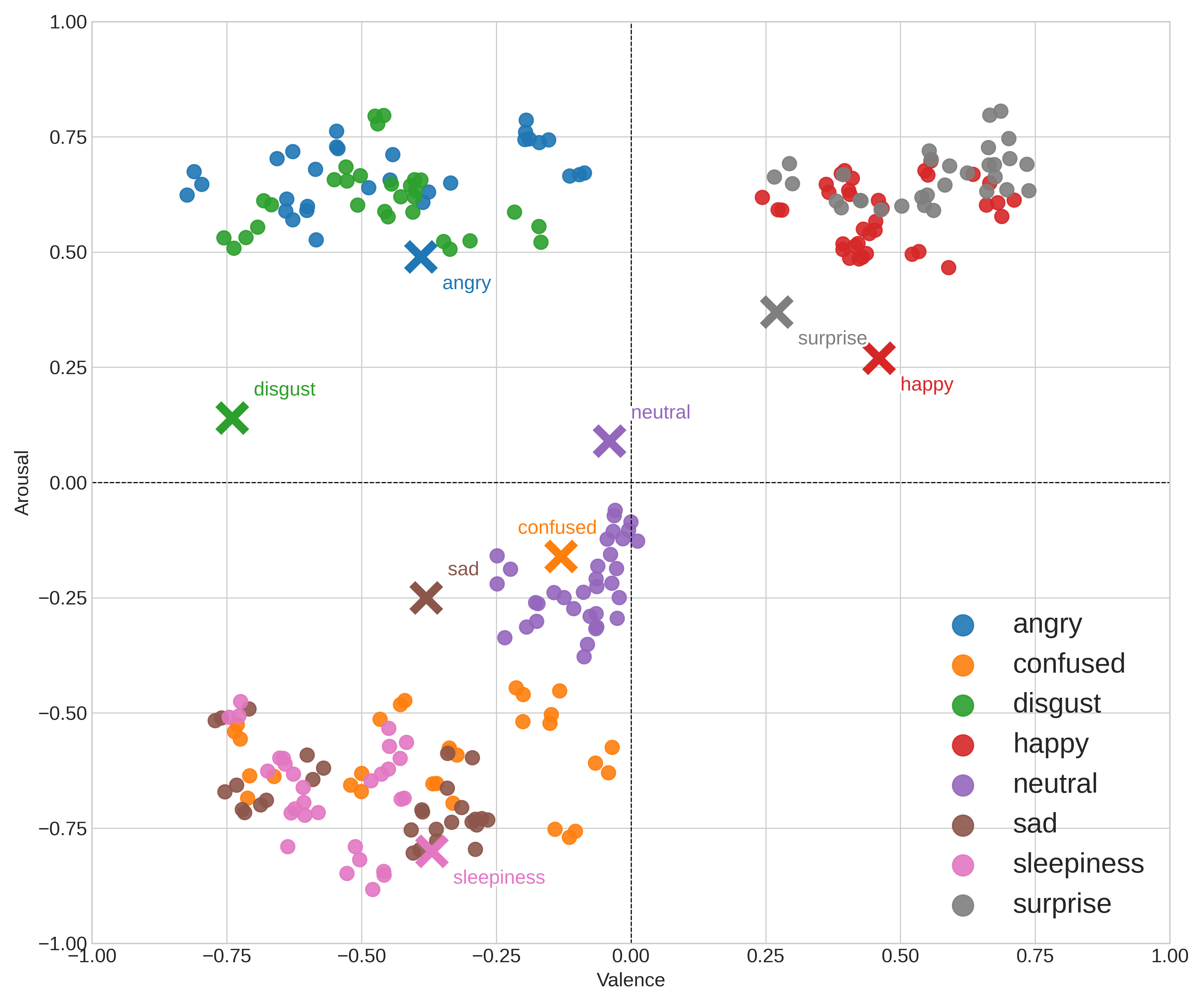}
        \subcaption{Chinese}
        \label{fig:china}
    \end{minipage}
    \hfill
    \begin{minipage}{0.33\textwidth}
        \centering
        \includegraphics[width=\textwidth]{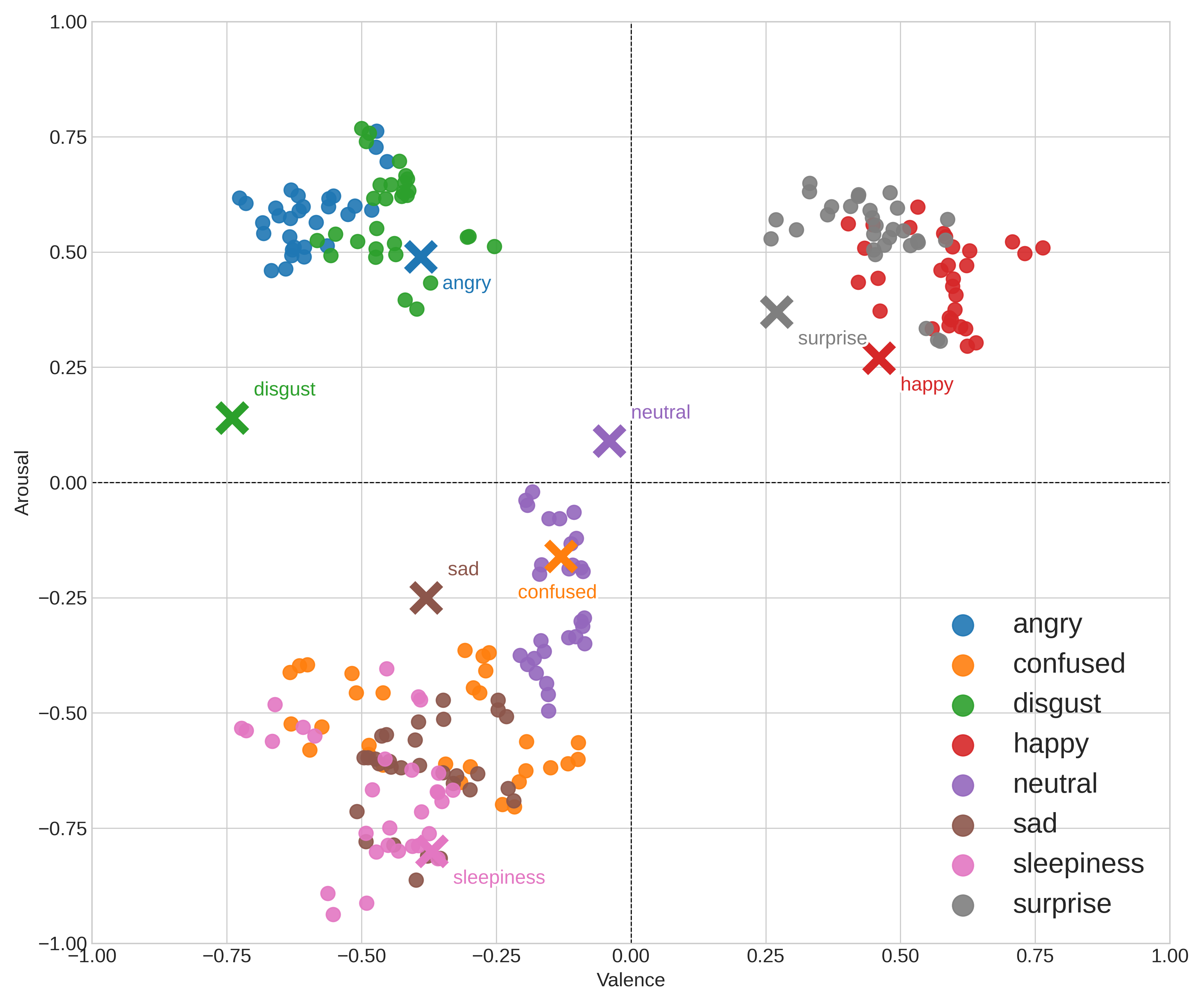}
        \subcaption{Indonesian}
        \label{fig:indo}
    \end{minipage}
    \hfill
    \begin{minipage}{0.33\textwidth}
        \centering
        \includegraphics[width=\textwidth]{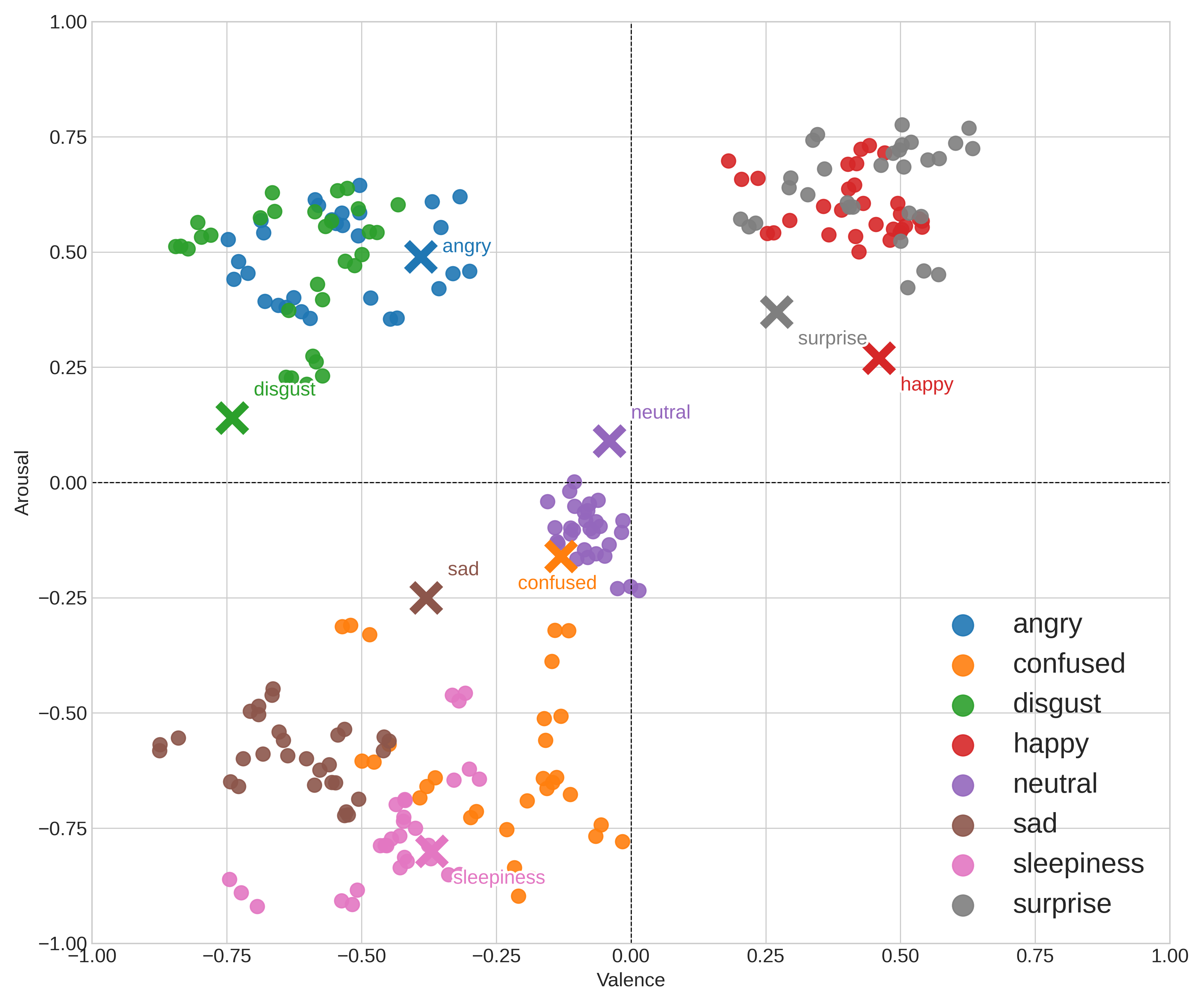}
        \subcaption{Japanese}
        \label{fig:jap}
    \end{minipage}
    \vspace{-0.3cm}
    \caption{A–V distributions from the personalization process for (a) Chinese, (b) Indonesian, and (c) Japanese participants. The $\bullet$ indicates participant-selected A–V values, and $\times$ indicates averaged reference dataset values.
    }
    \label{fig:culture}
\end{figure*}

\vspace{-0.1cm}
\section{Experimental Results}
\vspace{-0.1cm}

\subsection{Speech Quality and Emotional Similarity}
\vspace{-0.1cm}

Table~\ref{tab:mos} summarizes the performance of the emotional speech generator before human-in-the-loop IGA personalization. We compare the baseline Grad-TTS with emotion embedding against Grad-TTS with the proposed emotion controller.

Our model increases MOS from 3.37 to 3.75 and achieves a 23.5\% relative reduction in WER, indicating improved speech naturalness and intelligibility. Emotional similarity also improves, with CCC increasing from 0.60 to 0.84 for arousal and from 0.64 to 0.77 for valence, suggesting closer alignment between synthesized and reference emotional characteristics.

\begin{table}[t]
    \centering
    \vspace{-0.1cm}
    \caption{Results of MOS, WER, and CCC. Proposed method is shown in {\bf bold}. A: arousal, V: valence.}
    \vspace{-0.3cm}
    \label{tab:mos}
    \resizebox{\linewidth}{!}{%
    \begin{tabular}{lcccc}
        \toprule
          & \textbf{MOS} $\uparrow$ & \textbf{WER}(\%) $\downarrow$ & \textbf{CCC(A)} $\uparrow$ & \textbf{CCC(V)} $\uparrow$ \\
        \midrule
        Ground truth & 4.04 $\pm$ 0.09 & - & - & - \\
        \midrule
        Grad-TTS w/emo emb & 3.37 $\pm$ 0.10 & 21 & 0.60 & 0.64 \\
        \textbf{Grad-TTS w/emo ctrl} & \textbf{3.75 $\pm$ 0.09} & \textbf{17} & \textbf{0.84} & \textbf{0.77}\\
        \bottomrule
    \end{tabular}%
    }
    \vspace{-0.1cm}
\end{table}

% \vspace{-0.3cm}
\subsection{A–V Space after Personalization}
\vspace{-0.1cm}

Figure~\ref{fig:culture} shows A–V mappings obtained through the personalization process. Most personalized A–V values deviate from those in the original U.S.-based dataset, indicating that emotional perception varies across individuals and cultures.

Chinese participants (Fig.~\ref{fig:china}) exhibit relatively narrow arousal ranges but wider valence variation. Indonesian participants (Fig.~\ref{fig:indo}) show tighter clustering around emotion centers, suggesting more consistent emotional interpretation within the group. Japanese participants (Fig.~\ref{fig:jap}) tend to map negative emotions to lower arousal and require stronger expressions for positive emotions, reflecting a tendency toward restrained emotional expression.

\vspace{-0.1cm}
\subsection{Subjective Evaluation}
\vspace{-0.1cm}

\begin{figure}[t]
\centering
\vspace{-0.3cm}
\includegraphics[width=0.48\textwidth]{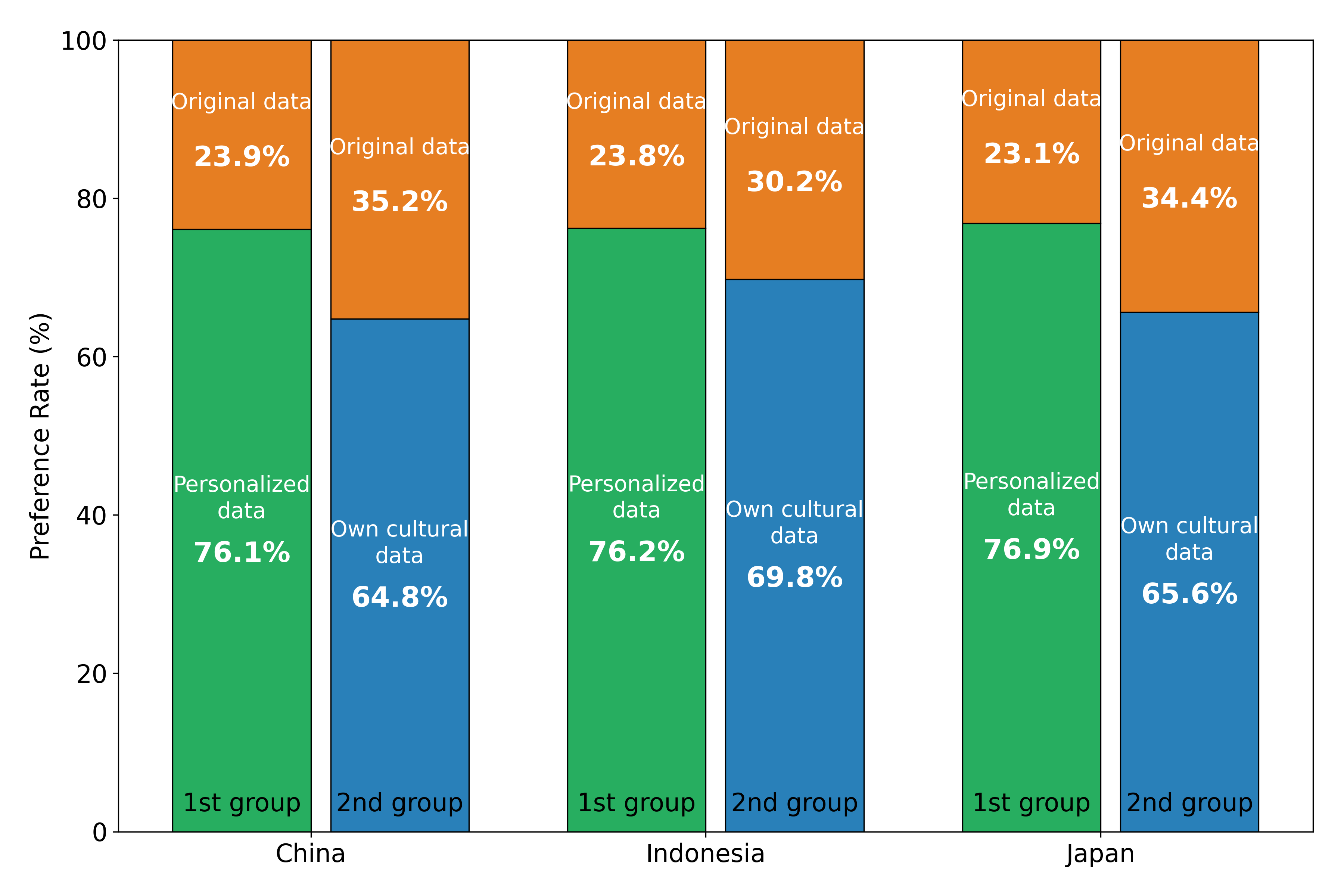}
\vspace{-0.8cm}
\caption{A/B preference test results.}
\label{fig:abtest}
\end{figure}

\textbf{Personalization:}
Participants who completed the IGA personalization process performed blind A/B comparisons between speech generated using their personalized A–V values and speech generated using the U.S.-based reference A–V baseline. As shown in Fig.~\ref{fig:abtest} (left bars), participants were unaware of which sample corresponded to their personalized mapping, yet personalized speech achieved a preference rate of 76\%. This result indicates that listeners consistently favored speech aligned with their own emotion perception.

\noindent\textbf{Culture-specific evaluation:}
A second group of participants who did not undergo personalization evaluated culturally adapted speech generated using culture-specific A–V averages (Chinese, Indonesian, and Japanese) against the U.S.-based baseline. As shown in Fig.~\ref{fig:abtest} (right bars), evaluations were conducted blindly, and participants were unaware of the cultural source of each sample. Preference rates were 64.8\%, 69.8\%, and 65.6\%, respectively, suggesting improved emotional alignment.

\noindent\textbf{Cross-cultural evaluation:}
In addition, we also conducted a cross-cultural evaluation in which participants ranked samples generated using Chinese, Indonesian, and Japanese A–V values. Participants from all three cultures consistently preferred samples generated with their own culture’s A–V mapping. Chinese participants selected Chinese samples 65\% of the time, while Japanese and Indonesian participants showed preferences of 67\% and 70\%, respectively. These preliminary results suggest group-level differences in preferred A-V mappings.

\vspace{-0.2cm}

% \vspace{-0.1cm}
\section{Conclusion}
\vspace{-0.1cm}
We presented a novel personalized emotional TTS framework that incorporates human-in-the-loop optimization with an Interactive Genetic Algorithm to address the subjectivity of emotion perception. Our results show that the proposed emotional controller improves both speech quality and emotional similarity, while personalization highlights individual and cultural variation in emotional preferences. These findings underscore the need to move beyond generic emotional models toward systems that adapt to both individual users and cultural contexts. Future work will explore broader linguistic and cultural settings, as well as strategies for real-time personalization, advancing the goal of emotionally adaptive conversational AI.

\newpage
\section{Acknowledgement} 
Part of this work was supported by JSPS KAKENHI Grant Numbers JP24K0296 and JP25H01139, as well as JST NEXUS (JPMJNX25C1).

\section{Generative AI Use Disclosure}
Generative AI tools were used only for minor language editing, such as grammar checking and wording refinement to improve clarity. All manuscript content and scientific work—including problem formulation, methodology development, experimental design and execution, and data analysis- were conducted entirely by the authors. The authors reviewed the final manuscript and take full responsibility for its accuracy and integrity.

\bibliographystyle{IEEEtran}
\bibliography{mybib}

\end{document}